# Conceptual design of a phase shifting telescope-interferometer

François Hénault

CRAL - Observatoire de Lyon, 9 Avenue Charles André, 69561 Saint Genis Laval, France

## Abstract

This paper deals with the theoretical principle and optical design of a phase-shifting telescope-interferometer. What is called a "Telescope-Interferometer" (T-I) is indeed a novel, recently proposed Wavefront Error (WFE) sensing technique, whose basic idea consists in combining the main pupil of a telescope with a second, off-axis reference arm. Then a weak modulation of the Point Spread Function (PSF) is generated at the focal plane, allowing for direct phase measurements. We propose a notable improvement of the method, inspired from classical principles of phase shifting interferometry. Herein are presented the alternative principle and its achievable measurement accuracy. The technique shows high performance excepted on narrow areas located near the pupil boundary. It is applicable to both ground or space telescopes and is suitable for the co-phasing of segmented mirrors, which is of prime importance in view of future giant telescope projects.

## 1   Introduction

Over the two last decades, progress in the field of Adaptive Optics (AO) for ground based telescopes led to outstanding improvements of astronomical observations. Nowadays this technology allows for routine study of numerous types of sky-objects in different spectral domains. However the performance of AO systems must continually be increased since new scientific challenges have recently emerged, e.g. the need for wider Fields of View (Multi-Conjugate or Ground-Layer AO) or the required extreme contrast ratios for extrasolar planets finding instruments. But the next-generation AO systems will also be faced to another technical challenge, which is the growing size of the future Extremely Large Telescope (ELT) projects. As a matter of fact these facilities will be equipped with segmented mirrors, like a few existing ground or space-borne telescopes such as the Keck 1 and 2, the Gran Telescopio Canarias or the James Webb Space Telescope (JWST). For such large-scale reflective surfaces, it is of prime importance to evaluate accurately the piston errors of each mirror segment, and to co-phase them within a reasonable lap of time. Unfortunately, the most familiar and efficient Wavefront Sensors (WFS) employed in Adaptive Optics are not well matched to that purpose, because they are all based upon WFE slopes or curvature measurements (then reconstructing the wavefronts digitally). Typical examples are the Shack-Hartmann [1], the shearing interferometer [2], the curvature sensor [3] and the most recently proposed pyramidal WFS [4]. Piston evaluation is hardly achievable from these devices, therefore direct phase measurement techniques should be preferred. It is worth mentioning, however, that recent studies demonstrated that the pyramidal WFS exhibits direct phase-sensing capacities when operated under certain conditions [5-6].

We should also notice that WFE retrieval – including piston defects – is already feasible by means of focal plane sensing algorithms: the latter were successfully employed to determine the optical aberrations of the Hubble Space Telescope [7-8], and represent the current basis for the co-phasing of the JWST hexagonal mirrors [9]. However such methods usually require extensive observing and post-processing times and are not appropriate to an on-ground adaptive optics regime, which has to correct high-frequency atmospheric disturbances.

Fast and direct phase measurement techniques have already been proposed by different authors. These WFS share the similar characteristic of being located at – or nearby – the telescope focal plane, such as the Mach-Zehnder interferometers described by Angel [10-11] and Codona [12] or holographic sensing suggested by Labeyrie [13]. However an alternative concept was recently proposed: this is the Telescope-Interferometer (T-I), which consists in combining the main pupil of a telescope with a second, off-axis reference arm [14]. Then a spatially modulated Point Spread Function (PSF) is generated in the image plane and measured by means of a detector array. It can be shown that the Optical Transfer Function (OTF) of the system (computed via inverse Fourier transform) provides quantitative information about the searched WFE of the telescope. Herein is proposed an alternative design of such a T-I, inspired from the well-known technique of Phase Shifting Interferometry (PSI) commonly used for optical surface metrology. Section 2 briefly summarizes the principle of a telescope-interferometer, mainly giving emphasis to the new concept. The theory is illustrated by numerical simulations presented in the chapter 3. Examples of practical implementation on a real telescope are described in section 4, while advantages and drawbacks of the proposed configuration are discussed in section 5. Finally, section 6 briefly concludes about future development of the method.

## 2 Principle of the phase-shifting telescope-interferometer

The general principle of the "off-axis" Telescope-Interferometer (T-I) was extensively described in Ref. [14]. Given a telescope of exit pupil diameter $D = 2R$, focal length $F$ (see Figure 1), and affected with an unknown wavefront error $\Delta(x,y)$, it consists in adding a second, de-centered optical aperture into the pupil plane. Then the light emitted by the same sky object is coherently combined to the principal beam and generates a spatially modulated PSF in the image plane. This weak amplitude fringe-pattern is demodulated by means of filtering in the Fourier plane, producing an OTF from which $\Delta(x,y)$ can be extracted. In the forthcoming sections, the telescope pupil will be named "main pupil" whereas the auxiliary optical arm is the "reference pupil". It must be stressed out that in any case the reference aperture must present two basic properties:
1) Its radius $r$ must be sensibly smaller than the telescope radius $R$, typically of one order of magnitude
2) It is assumed to be free of phase or alignment defects, thus generating a spherical reference WFE (this assumption will be revisited later in section 5).

The proposed concept of a phase shifting T-I is illustrated on the Figure 1. We suppose that the main pupil has an annular contour (a reasonable hypothesis when considering most of the existing ground or space telescopes) and that the reference pupil is re-centered on the Z optical axis of the telescope, then being concentric with the central obscuration. Hence the fringe demodulation technique for retrieving the original wavefront error $\Delta(x,y)$ is no longer feasible. Therefore some additional PSF intensity measurements must be performed, which may be realized for different consecutive phase shifts $\phi_0$ of the reference pupil. Following a similar approach than for the off-axis T-I, we shall first evaluate the PSF intensities measured in the focal plane, then define a phase retrieval algorithm from their associated OTFs.

## *2.1 Theoretical expression of the PSF intensities*

The considered coordinates systems are presented in Figure 1, where Z is the optical axis of the main telescope, OXY is the exit pupil plane, and O'X'Y' is attached to its focal plane, where the PSF intensities are being measured. The point O' corresponds to the telescope focus.

[Figure 1]

Let us denote $B_R(x,y)$ the two-dimensional amplitude transmission function of the main pupil, uniformly equal to 1 inside a circle of radius R, and to zero outside of this area. More generally this function may not have a circular contour (see for example the case of segmented mirrors) and shall also describe the central obscuration of the telescope, thus R should simply be considered as its clear aperture. If $\lambda$ is the wavelength of the incoming light (supposed to be monochromatic), the wave emerging from both main and reference apertures can be written in the OXY plane:

$$A_P(x, y) = B_R(x, y) \exp[i 2\pi \, \Delta(x, y)/\lambda] + B_r(x, y) \exp[i\phi_0] \quad (1)$$

where $B_r(x,y)$ is the amplitude transmitted by the reference pupil – here equal to the "pillbox" or "top-hat" function – and $\phi_0$ is the constant phase shift. In the frame of the Fraunhofer approximation, the wave diffracted in the O'X'Y' plane is obtained by Fourier transforming of $A_P(x,y)$:

$$A_P'(x', y') = FT[A_P(x, y)] = \iint_{x,y} A_P(x, y) \exp[-i 2\pi(ux + vy)] \, dx \, dy \quad (2)$$
$$= FT\{B_R(x, y) \exp[i 2\pi \, \Delta(x, y)/\lambda]\} + \exp[i\phi_0] \, FT[B_r(x, y)]$$

with $u = x'/\lambda F$ and $v = y'/\lambda F$. Here we shall slightly deviate from the classical formalism of Fourier optics, and normalize the diffracted wave with respect to the clear aperture of the optics, i.e. for the main pupil:

$$M(u, v) \exp[i \, \Psi(u, v)] = \frac{FT\{B_R(x, y) \exp[i 2\pi \, \Delta(x, y)/\lambda]\}}{\iint_{x,y} B_R(x, y) \, dx \, dy} = \frac{FT\{B_R(x, y) \exp[i 2\pi \, \Delta(x, y)/\lambda]\}}{S_R} \quad (3)$$

M(u,v) and Ψ(u,v) respectively are the modulus and phase terms of the complex amplitude diffracted in the telescope focal plane, and $S_R$ is the full collecting area of the main pupil. It can be noticed that both M(u,v) and Ψ(u,v) are directly linked to the WFE Δ(x,y) of the incident beam, on one hand, and that $M^2$(u,v) is proportional to the modulus of the PSF generated by the main aperture, on the other hand. Likewise, the wave propagated from the reference pupil can be rewritten as follows:

$$m(u, v) = \frac{FT\{B_r(x, y)\}}{\iint_{x,y} B_r(x, y)\, dx\, dy} = \frac{FT\{B_r(x, y)\}}{S_r} \qquad (4)$$

where m(u,v) is the square root of the modulus of the reference pupil PSF (in principle equal to the Airy disk), and $S_r = \pi r^2$ is the reference aperture area. Hence when combining relations (2) to (4), the expression of the complex amplitude $A_P'(x',y')$ becomes:

$$A_P'(x', y') = S_R \{ M(u, v) \exp[i\, \Psi(u, v)] + C \exp[i\phi_0] m(u, v) \} \qquad (5)$$

and C is a real number standing for the contrast ratio of the phase-shifting T-I:

$$C = S_r / S_R \qquad (6)$$

The intensity distribution $I_P'(x',y')$ in the focal plane is by definition equal to the square modulus of $A_P'(x',y')$, i.e. after multiplying with its complex conjugate:

$$I_P'(x', y') = S_R^2 \{ M^2(u, v) + C^2 m^2(u, v) \\ + C\, M(u, v)\, m(u, v)\, [\exp(i\, \Psi(u, v) - i\phi_0) + \exp(-i\, \Psi(u, v) + i\phi_0)] \} \qquad (7)$$

Once the PSF intensity measured on the telescope focal plane is known, we shall first compute the associated OTF, and then define a phase retrieval procedure.

## 2.2 Phase retrieval procedure

Most generally the OTF is expressed in terms of the pupil spatial frequencies $f_x$ and $f_y$ where $f_x$ = x/λF and $f_y$ = y/λF, and is limited by the cut-off frequency $f_c$ = 2R/λF above which it is uniformly equal to zero. Here the OTF must be re-scaled to the pupil plane coordinates x and y. This is achieved by simply ignoring the initial variable substitutions in the focal plane (i.e. u = x'/λF and v = y'/λF). Hence from Eqs. (3-4) and (6-7), and referring to some classical properties of Fourier optics theory regarding convolution products and Fourier transforms of complex conjugates, the phase-shifted OTF can be expressed as:

$$C_P(x, y) = FT^{-1}[I_P'(x', y')] \\ = S_R \left\{ O(x, y) + C \left( B_R(x, y) \exp\left[i\, \frac{2\pi}{\lambda} \Delta(x, y) - i\phi_0\right] + B_R(-x, -y) \exp\left[-i\, \frac{2\pi}{\lambda} \Delta(-x, -y) + i\phi_0\right] \right) \otimes \frac{B_r(x, y)}{S_r} \right\} \qquad (8)$$

where the symbol ⊗ denotes a convolution product, and the function O(x,y) stands for:

$$O(x, y) = O_R(x, y) + C\, O_r(x, y) \quad (9)$$

Here $O_R(x,y)$ and $O_r(x,y)$ respectively are the auto-correlation functions of the main and reference pupils, which are other classical definitions of the OTFs. These functions are extensively described in Ref. [15], section 9.5, and are always normalized such that their maximal values (at $x = y = 0$) are equal to 1. It may seem at first glance that phase extraction is hardly achievable from Eq. (8). However, it involves a crossed convolution product between both the main and reference pupils, whose phase is related to the searched wave-front error $\Delta(x,y)$. Hence the method shall be efficient if several PSF and OTF measurements (corresponding to different phase-shifts $\phi_0$) can be realized. This implies that four different data-set need to be acquired, for which $\phi_0$ will successively be equal to 0, $\pi/2$, $\pi$ and $3\pi/2$, thus leading to the following OTF expressions:

$$C_0(x, y) = S_R \left\{ O(x, y) + C \left( B_R(x, y) \exp\left[i \frac{2\pi}{\lambda} \Delta(x, y)\right] + B_R(-x, -y) \exp\left[-i \frac{2\pi}{\lambda} \Delta(-x, -y)\right] \right) \otimes \frac{B_r(x, y)}{S_r} \right\} \quad (10)$$

$$C_{\pi/2}(x, y) = S_R \left\{ O(x, y) - i\, C \left( B_R(x, y) \exp\left[i \frac{2\pi}{\lambda} \Delta(x, y)\right] - B_R(-x, -y) \exp\left[-i \frac{2\pi}{\lambda} \Delta(-x, -y)\right] \right) \otimes \frac{B_r(x, y)}{S_r} \right\} \quad (11)$$

$$C_{\pi}(x, y) = S_R \left\{ O(x, y) - C \left( B_R(x, y) \exp\left[i \frac{2\pi}{\lambda} \Delta(x, y)\right] + B_R(-x, -y) \exp\left[-i \frac{2\pi}{\lambda} \Delta(-x, -y)\right] \right) \otimes \frac{B_r(x, y)}{S_r} \right\} \quad (12)$$

$$C_{3\pi/2}(x, y) = S_R \left\{ O(x, y) + i\, C \left( B_R(x, y) \exp\left[i \frac{2\pi}{\lambda} \Delta(x, y)\right] - B_R(-x, -y) \exp\left[-i \frac{2\pi}{\lambda} \Delta(-x, -y)\right] \right) \otimes \frac{B_r(x, y)}{S_r} \right\} \quad (13)$$

and a phase retrieval formula can easily be derived from a simple linear combination of the four previous relationships:

$$\left( B_R(x, y) \exp\left[i \frac{2\pi}{\lambda} \Delta(x, y)\right] \right) \otimes \frac{B_r(x, y)}{S_r} = \frac{C_0(x, y) + i\, C_{\pi/2}(x, y) - C_{\pi}(x, y) - i\, C_{3\pi/2}(x, y)}{4\, C\, S_R} \quad (14)$$

The last result will appear at the same time strange and familiar to PSI users: familiar because it looks quite similar to their usual interferogram combination formulae, and strange since the complex number $i = (-1)^{1/2}$ is involved. In Figure 3 are provided three-dimensional views of the modules of one acquired OTF and of their linear combination, obtained for the segmented-mirror example described in section 3.

Eq. (14) demonstrates that the linearly combined OTFs are proportional to the convolution product of the complex amplitude transmitted by the main pupil with the circular pillbox function $B_r(x,y)$. Here a deconvolution process could be implemented, but might be detrimental to quasi real time AO applications. An alternative solution is to assume that the term $B_r(x,y)/S_r$ can be replaced with the Dirac distribution $\delta(x,y)$, thus neglecting the convolution product (this approximation is all the more valid as r/R tends to decrease). Consequently it may be assumed that:

$$B_R(x,y) \exp\left[i\frac{2\pi}{\lambda}\Delta(x,y)\right] \approx \frac{C_0(x,y) + i\,C_{\pi/2}(x,y) - C_\pi(x,y) - i\,C_{3\pi/2}(x,y)}{4\,C\,S_R} \quad (15)$$

and the WFE retrieval formula will finally be:

$$\Delta(x,y) \approx \frac{\lambda}{2\pi}\,\mathrm{Arctan}\left\{\frac{\mathrm{Im}[C_0(x,y) + i\,C_{\pi/2}(x,y) - C_\pi(x,y) - i\,C_{3\pi/2}(x,y)]}{\mathrm{Real}[C_0(x,y) + i\,C_{\pi/2}(x,y) - C_\pi(x,y) - i\,C_{3\pi/2}(x,y)]}\right\} \quad \mathrm{mod}\,[\lambda] \quad (16)$$

where Real[ ] and Im[ ] respectively stand for the real and imaginary parts of a complex number. Hence the basic theory and WFE reconstruction relationships of the phase shifting T-I have been defined. It must be noted however that due to the "convolution approximation" of Eq. (15), the absolute accuracy of the method will suffer from a systematic or "bias" measurement error. This effect was already studied in the case of the off-axis T-I and found to be acceptable [14]. Since the principle of the phase-shifting telescope is somewhat different however, we present in the following section an example of numerical simulation illustrating the achievable accuracy.

## 3    Computer simulation

In order to validate the global WFE retrieval procedure and to estimate the bias measurement error, we developed a simulation tool applicable to both types of T-I. For the phase-shifting telescope, the main steps of the numerical model are detailed on the flow-chart presented in Figure 2. As an illustration, Figure 3 shows the initial transmission map of the telescope-interferometer (a), the added WFE on the main pupil (d), the modulus (b) of a single OTF associated to a given phase-shift $\phi_0$, the modulus (c) of the four combined OTFs from Eq. (15), the retrieved WFE on the main pupil (e), and its difference map with respect to the initial WFE (f). Figs. 3-c and 3-e clearly demonstrate that both the amplitude (or intensity) and phase maps in the main pupil can be reconstructed by means of a phase shifting T-I.

[Figure 2]

In Ref. [14], a similar numerical model was applied to the case of an off-axis T-I, considering four different types of wave-front errors, which are pure defocus, low spatial frequency defects such as coma or astigmatism, random errors engendered by wave propagation into turbulent atmospheric layers, and the particular case of segmented mirrors that is of prime interest here. Table 1 summarizes the numerical results obtained in the latter case, for both types of telescope-interferometers. We considered a main telescope of exit pupil diameter 2R = 500 mm and focal length F = 5 m, thus having an effective aperture number of 10. The input monochromatic wavelength is assumed to be $\lambda$ = 1 µm. The diameter of the reference telescope is equal to 2r = 25 mm, hence the ratio between both apertures is 20 – a favorable condition to keep a good spatial resolution near the edges of the reflective segments. When turning clockwise from the Y-axis, the introduced piston errors present successive values of $\lambda/2$, $\lambda/4$, $-\lambda/4$, $-\lambda/2$, $-\lambda/4$ and $\lambda/4$, as illustrated in Fig 3-d where grey-levels are linearly scaled to the Peak-to-Valley (PTV) figures. Hence it can be noted that the intrinsic measurement accuracy of the phase-shifting T-I is approximately $\lambda/6$ PTV and $\lambda/110$ RMS. Such results are consistent with those previously obtained for the off-axis T-I (see Table 1), and are well under the diffraction limit of the main telescope according to the Rayleigh criterion. Using the numerical model, we also verified that comparable performance is still achieved for other typical WFE shapes (pure defocus, low spatial frequency and random defects).

[Table 1]

[Figure 3]

Although the attained measurement accuracy already looks quite satisfactory, it must be highlighted that most of the intrinsic error is due to the "convolution approximation" in Eq. (15), because the convolution product between the main and reference pupils tends to average useless information found outside the mirror segments with relevant data located inside them. Two alternative options can be suggested in order to enhance the performance of the method.

1) As the most significant errors are positioned near the main pupil rim (and more precisely over a width corresponding to the radius r of the reference pupil, see Fig. 3-f), the corrugated zone could be simply eliminated from the computing area. This might be achieved by multiplying the retrieved WFE with a pre-computed "pupil mask" at the end of the numerical procedure. Such a fast and easy technique may prove very efficient when only low order defects of the segmented mirrors are searched for (i.e. piston, tilt, defocus, etc.).

2) Another solution would be to directly solve Eq. (14) through a deconvolution process. This last idea looks reasonable since we already have an a priori knowledge of the convolution operand $B_r(x,y)/S_r$. However such algorithms are reputed for extensive computing times and remain quite infrequent for an AO operation. We finally consider that they should be the scope of a dedicated and independent study, thus bringing them beyond the scope of this paper.

Having asserted the general principle of a phase shifting telescope-interferometer, then attempted to evaluate its ultimate capacity with numerical simulations, we should then provide a more concrete description of the device and discuss its potential limitations. This is the purpose of the following sections.

# 4   Practical realization

## 4.1   Basic opto-mechanical design

Figure 4 provides a schematic description of one candidate optical design for the phase-shifting T-I (it must be emphasized that the lateral dimensions of the reference beam are stretched out for the sake of illustration). It shows a telescope from the Cassegrain family composed of a primary mirror (M1) and a secondary mirror (M2). Here the major difficulty practically consists in adding a reference optical arm centered on the optical axis of the main pupil, while maintaining an equal Optical Path Difference (OPD) between them. Among various conceivable configurations, we chose to inject the reference beam from behind the M2, through an opening located inside the central shadowed area. Then the reference telescope is split into two separated opto-mechanical assemblies. First, an afocal telescope formed by two confocal parabolic mirrors (M1' and M2') collects the light emitted by a sky-object already observed by the main pupil, and reflects the beam toward a second assembly, namely the "reference focusing telescope". The latter incorporates a Delay Line (DL), a folding mirror and a telephoto lens system of focal length f. F and f are linked by the relation F = f $M_r$, where $M_r$ is the angular magnification ratio of the afocal telescope. Several options can be envisaged to generate variable phase-shifts $\phi_0$ within the reference pupil arm. Three of them are schematically indicated in Figure 4, which are:
a) a classical Delay Line, appearing on the upper part of the reference focusing telescope assembly,
b) a set of variable thickness phase plates, here sketched on the exit port of the afocal telescope, but that could be located anywhere else into the parallel beam,
c) or an axial shift of the M2' mirror: as the required displacements are lower or equal than half a wavelength, it is likely that the small amount of resulting defocus remains negligible with respect to the systematic measurement errors.

Obviously the previous configuration should be optimized with regard to structural stability. It can be assumed that all the opto-mechanical units located at the immediate proximity of the M2 will benefit of an intrinsically stiff mounting structure. However the co-alignment of the reference afocal telescope with the main pupil might be more difficult to preserve, and could be assisted by means of a tip-tilt mirror replacing one of the two folding mirrors depicted in Figure 4.

[Figure 4]

### 4.2 Alternative designs

In the previous section was described a general layout for the implementation of a phase-shifting telescope-interferometer into an already existing facility (i.e. the main telescope). However, the previous concept could be greatly simplified if it is specified that the system should offer a phase-shifting capacity right from the start of the telescope design phase. In that case the needed reference pupil could be directly integrated on the optical surface of the giant primary mirror: it may consist in a piezoelectrically driven reflective facet that would successively be shifted of 0, $\pi/2$, $\pi$ and $3\pi/2$. Figure 5 shows different possible arrangements of the reference facet on the primary mirror, for both segmented and diluted-aperture geometry. The reader will easily verify that the whole set of equations (1-14) leading to the final phase retrieval formulae remains valid when the reference pupil is decentred of a given amount with respect to the main pupil: in that case Eq. [16] will simply generate a laterally-shifted pupil map without any information loss.

[Figure 5]

Another way to avoid the construction of the auxiliary telescope would be to give up a fraction of the light collected by the main pupil, injecting it into a mono-mode optical fiber that would spatially filter the reference beam and finally mix it with the main telescope beam along the optical axis. The device will then become quite similar to the "Michelson configuration" described in ref. [14], section 4, probably exhibiting the same kinds of advantages and drawbacks.

# 5  Discussion

The preceding sections essentially focused on the theoretical principle and ultimate performance of the phase shifting T-I, opening the way to some possible conceptual optical designs. However some fundamental issues remain to be discussed in order to state precisely the practical useful domain of the method. In particular the effects of some instrumental errors (image quality of the reference optical arm, vibrations, noise and non-linearity of the detectors) will naturally tend to worsen the ideal measurement accuracy predicted in section 3. Additionally, other error sources specific to PSI such as bias, drift or random errors of the phase-shift $\phi_0$ will also need to be addressed (in the latter case however, several corrective algorithms were already suggested and validated by different authors [16-17], and are most probably suitable to phase-shifting T-Is as well). A complete and quantitative analysis of the major measurement errors affecting a telescope-interferometer is already under preparation, and should be the scope of a future paper. However a few important points – that were already discussed in the case of the off-axis T-I – are commented hereafter, taking into account the peculiarity of the new on-axis, phase-shifting version.

   Signal-to-Noise Ratio (SNR) of the device: it readily appears that photometry and noise issues probably are the most critical areas of the method, at least in an AO operating mode, because it can be felt intuitively that the SNR is proportional to the contrast ratio C defined by Eq. (6). Then, from our first assumption in section 2, C should be much smaller than one (typically by two orders of magnitude), and the T-I may thus be particularly sensitive to photon noise. Additionally, the SNR issue is all the more critical that the method is limited by two fundamental constraints:
1) The observed sky-object must present a narrow spectral bandwidth $\Delta\lambda$, because the PSF of the T-I is scaled to the incident wavelength $\lambda$, while the inverse Fourier transform in Eq. (8) must be computed for a given reference wavelength $\lambda_0$. This will give rise to a scrambling of the measured PSFs and OTFs if $\Delta\lambda/\lambda_0$ exceeds a few percents.
2) The sky-object should not be spatially resolved by the main pupil in order to avoid contrast loss due to spatial incoherence, i.e. its angular radius shall not exceed $\lambda/2R$ typically.

When combining all the previous requirements (i.e. small contrast ratio, narrow spectral range and limited angular size of the luminous source), one may conclude that the main instrumental errors will originate from photon and read-out noise of the detector. However it was shown in ref. [14] that in first approximation the SNR of a T-I is not directly proportional to the contrast ratio, but rather to the effective collecting area $S_r$ of the reference telescope. Hence when considering large-scale facilities such as the future ELTs, the SNR should naturally increase with the diameter of the main pupil. This is an unusual behaviour for a WFS device, which obviously needs to be confirmed by further study.

   Image quality of the reference optical arm (also including the effects of vibrations): usually, PSI applied to optical surfaces metrology requires extremely tight polishing and manufacturing tolerances of the reference arm, in order to generate a quasi perfect WFE. However the phase-shifting T-I does not act in a similar way, as its intrinsic accuracy is limited by the convolution approximation of Eq. (15). To illustrate this, our numerical model showed that for the segmented mirror considered in section 3, WFEs of $\lambda/4$ PTV and misalignments of 0.5 arcsec have negligible influence (i.e. lower than half the intrinsic measurement error). Hence the reference telescope must not be completely free of phase or alignment defects as initially stated in § 2, but shall only be diffraction-limited, e.g. following the Rayleigh criterion. For that purpose, it might be equipped with its own AO system. It must be noted that similar conclusions were drawn for the off-axis T-I.

<u>Comparison between off-axis and phase-shifting T-I</u>: initiating a detailed discussion about the advantages and drawbacks of the two different T-I arrangements is probably too early, since the SNR and photometric performance of both devices are not yet assessed. However, some general trends may already be mentioned.

1) The main advantage of the off-axis T-I is the fact that it only requires a "single-shot" exposure, thus being insensitive to telescope instabilities – provided that the integration time can be made short enough. Conversely, the phase-shifting T-I requires four consecutive acquisitions, which imply improved opto-mechanical stability and may reduce the allowed measurement time, thus increasing detector noise.

2) But this apparent advantage may be balanced with the required angular size of the observed sky-objects: for an off-axis T-I, the latter shall not exceed the resolving power of the equivalent stellar interferometer [14], which is at most $\lambda/(3R+r)$, although a value of $\lambda/4R$ would be preferable. It leads to a gain around two on the apparent size of the observed star, and thus to a factor 1.5 in magnitude. Practically, this effect might finally benefit to the phase-shifting T-I.

3) From a general point of view, the manufacturing of both types of T-I might be considered of the same level of difficulty. However, when all the optional designs presented in this paper and in ref. [14] are considered, the more attractive undoubtedly is the phase-shifting T-I incorporating a reference facet into the primary mirror (see § 4.2).

<u>Comparison with other focal plane WFS</u>: though very exciting, trade-offs between the telescope-interferometer technique and focal plane WFS concepts such as depicted by Angel [11], Codona [12] or Labeyrie [13] cannot be undertaken in the frame of this study. In general, T-Is present the obvious drawbacks of requiring more complex structures (auxiliary telescope, delay line…) and of achieving weak contrast ratios. On the other hand, they need little additional optics near the focal plane, which could be used for simultaneous wavefront sensing and science exposures. Hence their low contrast is somewhat balanced by a higher throughput. In addition, it must be emphasised that T-I phase retrieval relationships (15-16) remain valid whatever is the amplitude of the considered phase defect (in fractions of $2\pi$). Conversely, the Mach-Zehnder focal plane interferometer [11] is limited to weak amplitude and phase errors, thus requiring a companion WFS system (taken from another family) in order to operate efficiently.

Finally, in non-AO operation, the method could rather be considered as an image-based sensing technique, where exposure times can typically be much longer. Hence the SNR shall be improved, but the telescope will have to fulfill more stringent stability requirements (that could probably be met for space-borne instruments only).

# 6  Conclusion

In this paper was discussed the theory and conceptual design of a phase shifting Telescope-Interferometer (T-I): this is a recently proposed wavefront sensing technique, which consists in adding a second, reference optical arm to the main pupil in order to generate modulated or phase-shifted PSFs and OTFs. The searched phase errors can then be extracted from the OTFs using relatively simple mathematical relationships. Numerical simulations demonstrated that direct phase retrieval is feasible, and that the achievable accuracy is at least as good as for the off-axis T-I. Then a schematic optical design was proposed and the advantages and drawbacks of the method were briefly discussed.

What is the real potential of these T-I techniques ? In theory they can realize fast and simultaneous co-phasing measurements of large segmented mirrors, which might be of key importance in view of future projects such as the ground-based ELTs or the space-borne Terrestrial Planet Finder (TPF). However it is clear that their ultimate performance closely depends on photometry issues (see section 5), and that only an in-depth analysis of related topics (including noises of the CCD detector, useful spectral range and angular size of the observed stars) can bring a firm answer. This forthcoming study implies a notable development of our numerical model. It will also be helpful to define the best operating mode of the device that may either be suited to an Adaptive Optics regime, or limited to periodical diagnostics of the telescope figure, just like image-based sensing methods already are.

# References

[1]  R.G. Lane and M. Tallon, "Wave-front reconstruction using a Shack-Hartmann sensor," Appl. Opt. vol. 31, p. 6902-6908 (1992).

[2]  J.W. Hardy, J.E. Lefebvre and C.L. Koliopoulos, "Real-time atmospheric compensation," J. Opt. Soc. Am. vol. 67, p. 360-369 (1977).

[3]  F. Roddier, "Curvature sensing and compensation: a new concept in adaptive optics," Appl. Opt. vol. 27, p. 1223-1225 (1988).

[4]  R. Ragazzoni, "Pupil plane wavefront sensing with an oscillating prism," Journal of Modern Optics vol. 43, p. 289-293 (1996).

[5]  C. Vérinaud, "On the nature of the measurement provided by a pyramidal wave-front sensor," Optics Communications vol. 233, p. 27-38 (2004).

[6]  O. Guyon, "Limits of adaptive optics for high-contrast imaging," Astrophysical Journal vol. 629, p. 592-614 (2005).

[7]  J.R. Fienup, J.C. Marron, T.J. Schulz and J.H. Seldin, "Hubble Space Telescope characterized by using phase-retrieval algorithms," Appl. Opt. vol. 32, p. 1747-1767 (1993).

[8]  C. Roddier and F. Roddier, "Combined approach to the Hubble Space Telescope wave-front distortion analysis," Appl. Opt. vol. 32, p. 2992-3008 (1993).

[9]  D. Redding, S. Basinger, D. Cohen, A. Lowman, F. Shi, P. Bely, C. Bowers, R. Burg, L. Burns, P. Davila, B. Dean, G. Mosier, T. Norton, P. Petrone, B. Perkins and M. Wilson, "Wavefront control for a segmented deployable space telescope," in *UV, Optical, and IR Space Telescopes and Instruments*, J.B. Breckinridge and P. Jakobsen eds., Proceedings of the SPIE vol. 4013, p. 546-558 (2000).

[10]  R. Angel, "Ground-based imaging of extrasolar planets using adaptive optics," Nature vol. 368, p. 203-207 (1994).

[11]  R. Angel, "Imaging extrasolar planets from the ground," in *Scientific Frontiers in Research on Extrasolar Planets*, D. Deming and S. Seager eds., ASP Conference Series vol. 294, p. 543-556 (2003).

[12]  J.L. Codona and R. Angel, "Imaging extrasolar planets by stellar halo suppression in separately corrected color bands," The Astrophysical Journal vol. 604, p. L117-L120 (2004).

[13]  A. Labeyrie, "Removal of coronagraphy residues with an adaptive hologram, for imaging exo-Earths," in Astronomy with High Contrast Imaging II, C. Aime and R. Soummer eds., EAS Publications Series vol. 12, p. 3-10 (2004).

[14]     F. Hénault, "Analysis of stellar interferometers as wavefront sensors," Appl. Opt. vol. 44, p. 4733-4744 (2005).

[15]     M. Born and E. Wolf, *Principles of optics* (London, Pergamon, 6th ed., 1980).

[16]     J. Schwider, R. Burow, K.E. Elssner, J. Grzanna, R. Spolaczyk and K. Merkel, "Digital wave-front measuring interferometry: some systematic error sources," Appl. Opt. vol. 22, p. 3421-3432 (1983).

[17]     K. Kinnstaetter, A.W. Lohmann, J. Schwider and N. Streibl, "Accuracy of phase shifting interferometry," Appl. Opt. vol. 27, p. 5082-5089 (1988).

## Tables Caption

Table 1: Comparison between off-axis and phase-shifting telescope-interferometers

**Table 1: Comparison between off-axis and phase-shifting telescope-interferometers**

|  |  | INITIAL WAVE-FRONT ERROR | RETRIEVED WAVE-FRONT ERROR | DIFFERENCE MAP | ERROR RATIO (%) |
|---|---|---|---|---|---|
| **Off-Axis T-I** | PTV ($\lambda$) | 0,997 | 0,997 | 0,182 | 18,2 |
|  | RMS ($\lambda$) | 0,324 | 0,323 | 0,011 | 3,5 |
| **Phase-Shifting T-I** | PTV ($\lambda$) | 0,997 | 0,997 | 0,166 | 16,7 |
|  | RMS ($\lambda$) | 0,350 | 0,349 | 0,009 | 2,5 |

# Figures Caption

Figure 1: Coordinate systems
Figure 2: Flow-chart of the numerical model
Figure 3: A typical example of WFE reconstruction sequence. Top row, from left to right: full transmission map of the T-I, including the central reference pupil (a), MTF derived from a single acquired OTF (b), and the four combined OTFs (c). Bottom row: initial (d) and reconstructed (e) WFEs of segmented mirror defects, and their bidimensional difference-map (f). Grey-levels are scaled to PTV values
Figure 4: Possible implementation on a real telescope
Figure 5: Integrating a reference pupil in giant primary mirrors. For segmented reflective surfaces, the reference facet can either be a piezoelectrically driven segment (a) or a different mirror installed in place of it (b). For diluted-aperture telescopes, several locations are possible (c), depending on the geometry of the main pupil

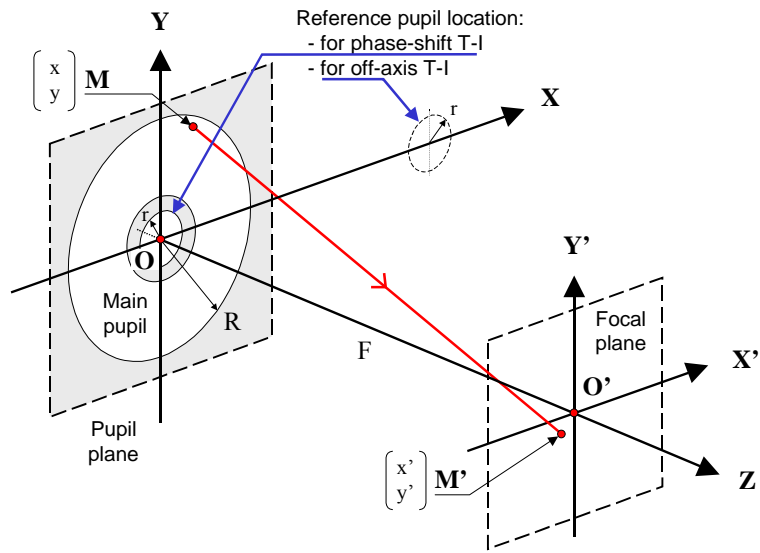

**Figure 1: Coordinate systems**

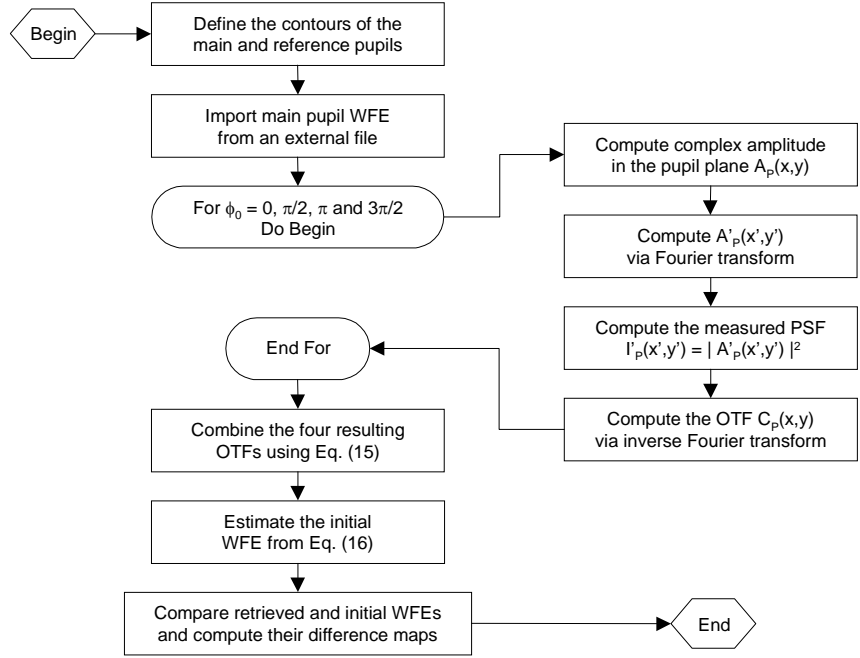

**Figure 2: Flow-chart of the numerical model**

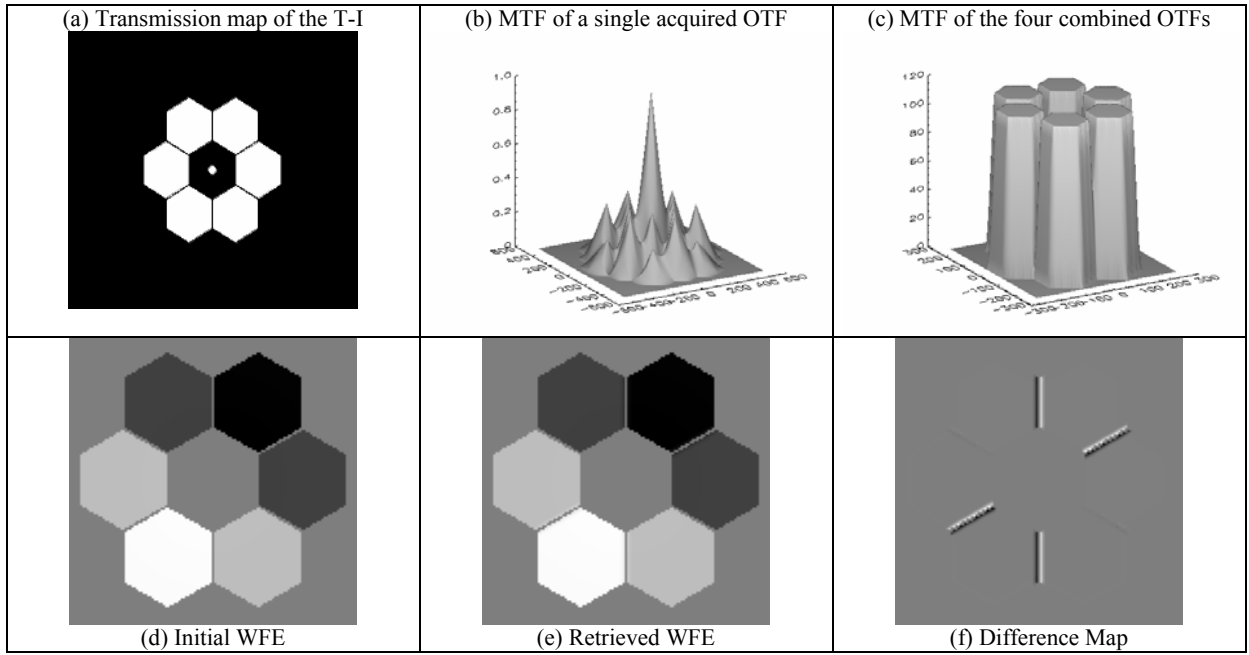

**Figure 3: A typical example of WFE reconstruction sequence. Top row, from left to right: full transmission map of the T-I, including the central reference pupil (a), MTF derived from a single acquired OTF (b), and the four combined OTFs (c). Bottom row: initial (d) and reconstructed (e) WFEs of segmented mirror defects, and their bidimensional difference-map (f). Grey-levels are scaled to PTV values**

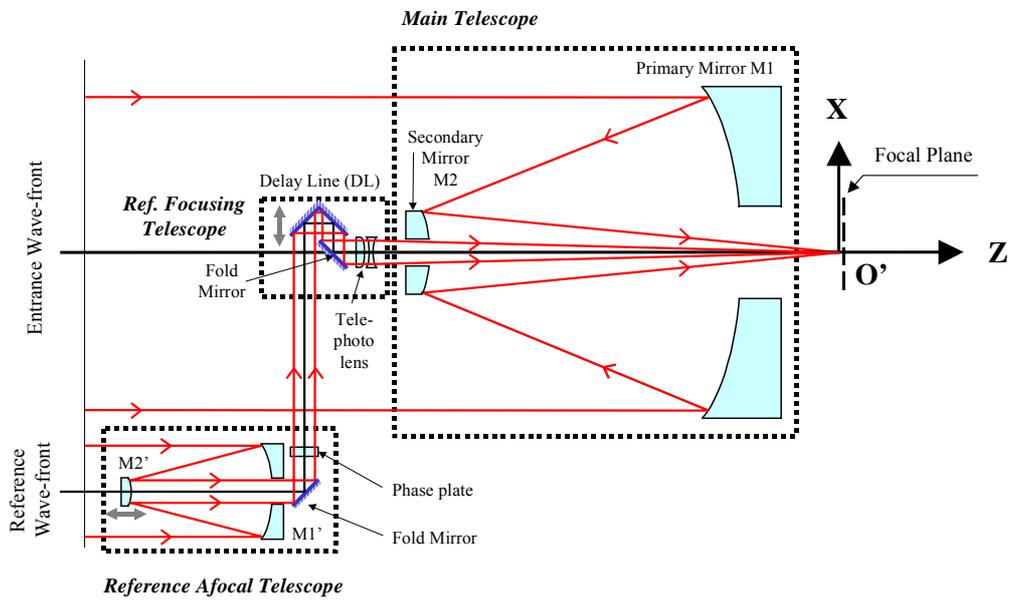

**Figure 4: Possible implementation on a real telescope**

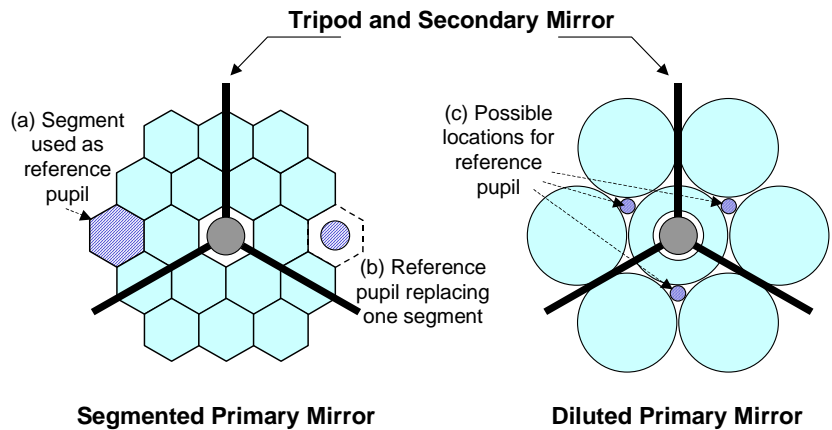

**Figure 5: Integrating a reference pupil in giant primary mirrors. For segmented reflective surfaces, the reference facet can either be a piezoelectrically driven segment (a) or a different mirror installed in place of it (b). For diluted-aperture telescopes, several locations are possible (c), depending on the geometry of the main pupil**

17